\documentstyle[twocolumn,aps,epsfig]{revtex}
\begin{document}
\draft
\title{Rotating Atomic Traps for Bosons and the Centrifuge Effect}
\author{T. Majmudar and A. Widom}
\address{Physics Department, Northeastern University, Boston MA 02115}
\maketitle

\begin{abstract}
A rigorous time independent Hamiltonian for rotating  atomic traps is 
discussed. The steady states carry a mass current and thereby an angular 
momentum. It is shown that the rotation positions the atoms away from the 
rotation axis (after taking both the time and quantum mechanical 
averages) as in a conventional centrifuge. Some assert that 
the rotation for Bose condensates cause the atoms to move towards 
the rotation axis; i.e. act oppositely to fluids in a centrifuge. 
The opposing physical pictures are reminiscent of the difference 
between the rotational motion views of Newton and Cassini. 
\end{abstract}  

\pacs{PACS: 05.30.J, 03.75.F, 03.75.F, 42.50.G}  
\narrowtext

\section{Introduction} 

In recent years there has been considerable interest in the possibility 
of Bose condensation\cite{1,2,3} of trapped mesoscopic clusters 
of atoms\cite{4}. One of the more difficult experimental problems to 
solve has been the design of effective traps\cite{5,6,7,8}. 
An often used Bose fluid container employs the ``time averaged 
orbiting potential'' (TOP) trap\cite{9}. 

\begin{figure}[htbp]
\begin{center}
\mbox{\epsfig{file=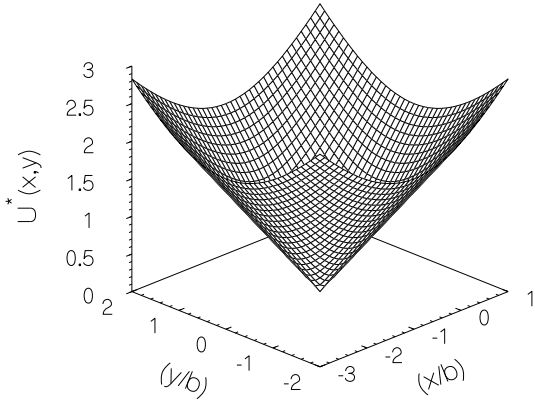,height=80mm}}
\caption{Shown is the conical potential produced by a static 
magnetic field  $U(\rho ,\phi ,z=0)= 
\hbar \gamma Gb U^*(\rho \cos \phi,\rho \sin \phi )$ as defined 
in Eq.(1).}
\label{topfig1}
\end{center}
\end{figure}

The design of the TOP trap begins with a magnetic bottle formed by the 
superposition of a uniform magnetic field and a quadrapole magnetic 
field, i.e. the effective single atom adiabatic potential in the 
magnetic bottle is given (in cylindrical coordinates) by 
\begin{equation}
U(\rho ,\phi ,z)=\hbar \gamma G 
\sqrt{\rho^2+b^2 +4z^2+2\rho b \cos \phi }
\end{equation}
where \begin{math} \gamma  \end{math} is the atomic gyromagnetic ratio, 
\begin{math} G  \end{math} the quadrapole magnetic field gradient 
and  \begin{math} b=(B_0/G) \end{math}, where  \begin{math} B_0  \end{math} 
is the magnitude of uniform part of the magnetic field. In the 
\begin{math} z=0  \end{math} plane, the conical potential is shown in Fig.1.

An experimental problem with the conical potential is that atoms 
located near the tip of the cone can make a transition onto another 
adiabatic potential and drift away. There is a leak at the bottom 
of the cone. In order to plug up the leak, one must move the 
atoms away from the bottom tip of the cone. In some experiments, 
a laser beam blasts the atoms away from the tip. In other 
experiments one simply rotates the potential around an axis as 
shown in FIG.2

\begin{figure}[htbp]
\begin{center}
\mbox{\epsfig{file=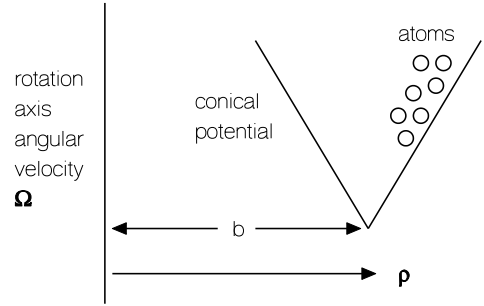,height=80mm}}
\caption{In order to move the atoms away 
from the tip of the conical potential, the full 
potential is rotated about an axis. Shown are some atoms 
moved away from the potential minimum by the rotation.}
\label{topfig2}
\end{center}
\end{figure}

One may try\cite{10,11} to treat the rotating potential in analogy 
with the workings of a centrifuge. It might be imagined 
that the atoms are thrown outwards to large values of 
\begin{math} \rho >b  \end{math} as shown schematically in the FIG.2.
However, it has been reported in the literature that Bose condensed 
atoms are thrown inwards to smaller values of the radial coordinates 
\begin{math} \rho <b  \end{math}\cite{9}. The observation of such a TOP 
trap contraction of the cluster has not been direct. Most of the direct 
observations of the atomic positions take place after the trap potential 
has been removed\cite{9,12}. However, the notion of the contraction of 
the cluster would follow theoretically if one replaced the actual potential 
in Eq.(1) with the time averaged (over one rotational period) potential
\cite{13,14,15,16,17,18,19,20} 
\begin{equation}
\bar{U}(\rho ,z)=\left({\Omega\over 2\pi}\right)
\int_{-\pi/\Omega }^{\pi/\Omega }  
U(\rho ,\phi +\Omega t,z)dt.
\end{equation}

Our work is organized as follows: In Sec.II a rigorous time 
independent Hamiltonian for a rotating potential will 
be derived\cite{21,22}. It will be shown that 
the time independent energy eigenstates carry a mass current 
\begin{math} {\bf J}  \end{math}, and thereby an angular momentum 
\begin{equation}
{\bf L}=\int ({\bf r\times J})d^3{\bf r}
=\sum_{i=1}^N ({\bf r}_i\times {\bf p}_i) .
\end{equation}
The rotating fluid with angular momentum 
\begin{math} {\bf L}  \end{math} is held inside the bottle with 
a force pointing towards the rotation axis as shown in FIG.2. 
In order for the force to point towards the rotation axis, the atoms 
must be positioned so that \begin{math} \rho>b  \end{math}. If 
\begin{math} \rho<b  \end{math}, then the bottle walls push the atoms 
away from the axis. It has been maintained in the literature on 
top traps that the atoms are positioned so that 
\begin{math} \rho<b  \end{math} which implies a wall force directed 
away from the rotation axis. In Sec.III, rigorous time averaging theorems 
will be proved. In the concluding Sec.IV, the 
experimental importance of the theoretical results will be examined. 

\section{Rotational Hamiltonian}

With the single atom Hamiltonian 
\begin{equation}
\tilde{h}_i(t)=-\left({\hbar ^2\over 2M}\right)\nabla_i^2+
U(\rho_i ,\phi_i+\Omega t ,z_i),
\end{equation}
the Hamiltonian for \begin{math} N \end{math} atoms in a TOP trap 
is given by 
\begin{equation}
H(t)=\sum_{i=1}^N \tilde{h}_i(t) +\sum_{i<j}^N u_{ij}
\end{equation}
wherein the two body potential commutes with the angular momentum 
operator  
\begin{equation}
{\bf L}=-i\hbar \sum_{i=1}^N {\bf r}_i{\bf \times \nabla }_i.
\end{equation}
Employing the canonical transformation
\begin{equation}
{\cal H}=S^\dagger (t)H(t)S(t)
-i\hbar S^\dagger (t){\partial S(t)\over \partial t}
\end{equation}
with  
\begin{equation}
S(t)=exp\left(-i\Omega L_z t/\hbar)\right),
\end{equation}
one finds a rigorously exact expression for the time independent 
Hamiltonian corresponding to a TOP angular velocity 
\begin{math} \Omega  \end{math}
\begin{equation}
{\cal H}=\sum_{i=1}^N h_i
+\sum_{i<j}^N u_{ij},
\end{equation}
where 
\begin{equation}
h_i=\tilde{h}_i(0)-\Omega l_i, \ \ \ 
l_i=-i\hbar \left({\partial \over \partial \phi_i}\right).
\end{equation}
In more detail, the {\em time independent} Hamiltonian for  
atoms in a TOP trap has the form 
$$
{\cal H}={1\over 2M}\sum_{i=1}^N ({\bf p}_i-M{\bf\Omega \times r}_i)^2
$$
\begin{equation}
+\sum_{i<j}^N u({\bf r}_i,{\bf r}_j),
-{M\over 2}\sum_{i=1}^N \left|{\bf \Omega \times r}_j\right|^2
\end{equation}
where \begin{math}{\bf \Omega } \end{math} is the angular velocity 
(axial) vector along the \begin{math} z \end{math}-axis. The last term 
on the right hand side of Eq.(11) represents the centrifugal (effective) 
potential energy.

That the Hamiltonian \begin{math} {\cal H} \end{math} describes the 
TOP trap atoms in a rotating frame is evident from the velocity 
operator for the \begin{math}i^{th}\end{math} atom.
\begin{equation}
{\bf v}_i=\left({i\over \hbar }\right)\left[{\cal H},{\bf r}_i\right]
=\left({{\bf p}_i\over M}\right)-{\bf \Omega \times }{\bf r}_i.
\end{equation}
Note that two components of the velocity operator do not commute; i.e. 
\begin{equation}
\left[v_{xi},v_{yj}\right]
=-\left({2i\hbar \Omega \over M}\right)\delta_{ij}.
\end{equation}
The acceleration 
\begin{equation}
{\bf a}_i=\left({i\over \hbar }\right)\left[{\cal H},{\bf v}_i\right]
\end{equation}
is given by 
\begin{equation}
M{\bf a}_i={\bf f}_i+M\left(2{\bf v}_i{\bf \times \Omega }+
{\bf \Omega \times}({\bf r}_i{\bf \times \Omega })\right).
\end{equation}
where the internal atomic forces and the confining force obey 
\begin{equation}
{\bf f}_i=-{\bf \nabla }_i\sum_{j\ne i}^N  u({\bf r}_i,{\bf r}_j)
-{\bf \nabla }_iU({\bf r}_i).
\end{equation}
In Eq.(16), the confining potential \begin{math} U({\bf r}) \end{math} 
is given in Eq.(1). Eq.(15) describes the normal, Coriolis and 
centrifugal forces in a fully quantum mechanical operator framework.

Summing Eqs.(15) and (16) over all of the atoms confined in the TOP trap,
yields 
$$
M\sum_{i=1}^N {\bf a}_i+\sum_{i=1}^N {\bf \nabla }_iU({\bf r}_i)=
$$
\begin{equation}
M\sum_{i=1}^N 
\left(2{\bf v}_i{\bf \times \Omega }+
{\bf \Omega \times}({\bf r}_i{\bf \times \Omega })\right),
\end{equation}
where the internal forces in the sum cancel due to the equality of 
action and reaction forces, i.e. momentum conservation. 

Thus far, our considerations are rigorously true for the model. The  
rigorously exact results can be extended to theorems regarding the 
quantum and also the time averaged properties of the system. The 
following two theorems are of central importance. 
\par \noindent 
{\bf Theorem I:} For a stationary state density matrix 
\begin{math} \rho \end{math} obeying 
\begin{math}i\hbar \dot{\rho }=\left[{\cal H},\rho \right]=0 \end{math}, 
the mean value of the time rate of change of a bounded quantity 
\begin{math} Q \end{math}
vanishes i.e.  
\begin{equation} 
\left<\dot{Q}\right>=0. 
\end{equation}
\par \noindent 
{\bf Proof:} Employing 
\begin{math} \dot{Q}=(i/\hbar )\left[{\cal H},Q \right] \end{math}
and the cyclic invariance of the trace
\begin{equation}
\left<\dot{Q}\right>=Tr\left(\rho \dot{Q}\right)
=Tr\left(
\dot{\rho }Q
\right)=0. 
\end{equation}
To apply this theorem for atoms in a rotating trap, let us consider 
mean acceleration of one atom via Eqs.(17); i.e. 
\begin{equation}
\left<\big({\bf a}+2{\bf \Omega \times v}
+{\bf \Omega \times }({\bf \Omega \times r})\big)\right>
=-\left<{\bf \nabla }U({\bf r})\right>/M.
\end{equation}
In a stationary state, Eq.(18), 
\begin{math}<{\bf a}>=<\dot{\bf v}>=0 \end{math} and 
\begin{math}<{\bf v}>=<\dot{\bf r}>=0\end{math}. 
Thus we arrive at 
\par \noindent 
{\bf Theorem II:} In any rotational stationary state, the (mean) 
mass times the centripetal acceleration of an atom is equal to 
the (mean) force exerted on the atom by the confining potential 
\begin{equation}
M\left<\big({\bf \Omega \times }({\bf \Omega \times r})\big)\right>
=-\left<{\bf \nabla }U({\bf r})\right>.
\end{equation}

Let us now return to FIG.2. Clearly the (mean) centripetal acceleration  
\begin{math} 
{\bf \Omega \times }({\bf \Omega \times}<{\bf r}>)  
\end{math}
points towards the rotation axis. Thus, the mean force due to the 
wall potential  
\begin{math}
(-<{\bf \nabla }U({\bf r})>)
\end{math}
also points towards the rotational axis. This proves, beyond any doubt, 
that the atoms must be positioned at a distances 
\begin{math} \rho >b \end{math}
as in FIG.2. It is {\em not possible} in a stationary state to have the atoms 
on the average at a distances closer than \begin{math} b \end{math}.

The above theorems can be further extended to non-stationary states if 
time averaging techniques are employed. If 
\begin{math} <Q(t)> \end{math} is a quantum mean value of a physical 
quantity at time \begin{math} t \end{math}, then the time average 
of that mean value is defined by 
\begin{equation}
\overline{Q}=\lim_{\tau \to \infty}\left({1\over \tau}\right)
\int_{t_0-(\tau /2)}^{t_0+(\tau /2)} \left<Q(t)\right> dt.
\end{equation}
It is a simple matter to prove the following 
\par \noindent 
{\bf Theorem III:} If \begin{math} <Q(t)>  \end{math} is a bounded function 
of time, then 
\begin{equation}
\overline{\dot{Q}}=0.
\end{equation}
\par \noindent 
{\bf Proof:} 
$$
\overline{\dot{Q}}=\lim_{\tau \to \infty}\left({1\over \tau}\right)
\int_{t_0-(\tau /2)}^{t_0+(\tau /2)} {d\left<Q(t)\right>\over dt} dt=
$$
\begin{equation}
\lim_{\tau \to \infty}
\left({<Q(t_0+(\tau /2))>-<Q(t_0-(\tau /2))>\over \tau}\right)=0.
\end{equation}
Finally, proceeding as before we prove the central result of this work.
\par \noindent 
{\bf Theorem IV:} In any state with finite quantum and time averages, 
the mass times the mean centripetal acceleration of an atom is equal to 
the mean force exerted on the atom by the confining potential; 
\begin{equation}
M{\bf \Omega \times }({\bf \Omega \times } \overline{{\bf r}})
=-\overline{{\bf \nabla }U}.
\end{equation}
{\bf Proof:} Apply Eq.(24), in the form 
\begin{math} \overline{\bf a}=0  \end{math} and 
\begin{math} \overline{\bf v}=0  \end{math}, to Eq.(20).

For a classical centrifuge with rotating walls, our central Eq.(25) 
can be employed to prove the usual result that particle are thrown 
outwards by the rotation. What we have quite rigorously proved is that 
Eq.(25) is also true for identical Bosons with quantum mechanical effects 
fully taken into account. 

\section{Conclusions}

The reported experimental studies which maintain 
that the rotating TOP trap pulls the particles inward (opposite to 
the centrifuge effect) appears more than just a little puzzling 
to us. As far as we know there have been no direct observations 
of particles localized on the rotation axis of a TOP trap. (i) In 
some cases the atoms have been observed flying outwards after the 
TOP trap has been removed. (ii) In an {\em in situ} 
measurement\cite{23}, the absorption of light by atoms 
in the trap was from an incident beam directed normal to the 
rotation axis. Again, the notion of atoms clustered on the 
rotation axis is (at best) only indirectly inferred from 
experimental data. (iii) In {\em in situ} measurements, the 
probe pulse of light is synchronized to the angular velocity 
of rotation. If the fluid in the trap responded  {\em 
only} to the time averaged potential, then synchronization should  
have no experimental consequences. Thus the notion of employing 
a static harmonic oscillator potential to model atoms in a 
dynamic TOP trap may be unreliable. 

Most of the central theorems proved above for rotating quantum 
mechanical systems, have been previously derived for classical 
fluids rotating in steady state. When the fluids are in a 
state of rotational flow, they tend to form ellipsoidal figures 
of equilibrium. The history of mathematical studies of the 
stability of such ellipsoidal figures has been reviewed in detail 
by Chandrasekhar.\cite{24} The averaging procedure for classical 
rotating fluids relies heavily on virial moments of the mass an 
velocity distributions. Some of the final results obtained by 
these classical virial methods are identical to those we have 
achieved by time and quantum mechanical averaging procedures.  
The classical rotational high angular velocity centrifuge 
effects retain their validity even in the quantum domain.

Finally, Newton\cite{25} derived the oblate spheroid shape for the 
rotating earth employing what is presently very well known as a 
centrifuge effect. The equatorial circle has a {\em slightly larger}  
diameter than distance between the north and south pole due to 
the daily rotation of the earth. Four generations of the Cassini 
family\cite{24,26} argued that the earth was a prolate spheroid 
with the equatorial circle having a {\em slightly smaller} 
diameter than distance between the north and south pole. The Cassini 
family argued (against Newton) that the earth's mass would be drawn 
towards the rotational axis of the earth, not unlike what has been 
claimed for atoms in a rotating TOP trap. The considerations of the Cassini 
family were shown to be incorrect. While the settlement of 
the shape of the earth required long times and large human and economic 
expenses of sizable expeditions spreading from France to Lapland to Peru, 
it is to be hoped that direct observations of fluid shapes in TOP 
traps can be carried out more expeditiously.

\end{document}